\documentclass[useAMS,usenatbib]{mnras}
\usepackage{graphicx}
\usepackage{color}
\usepackage{amssymb}
\usepackage{amsmath}
\usepackage{subfigure}
\usepackage{rotating}
\usepackage{pdflscape}
\usepackage{bm} 
\usepackage{mathtools}

\graphicspath{{images/}}

\title[{Galactic cosmic ray intensities for nearby stars}]{{Charting nearby stellar systems: The intensity of Galactic cosmic rays for a sample of solar-type stars}}
\author[Rodgers-Lee, Vidotto, Mesquita]{D. Rodgers-Lee$^{1}$\thanks{E-mail:
drodgers@tcd.ie}, A. A. Vidotto$^{1}$ and A. L. Mesquita$^{1}$,  \\ 
$^{1}$ School of Physics, Trinity College Dublin, University of Dublin, College Green, Dublin 2, D02 PN40, Ireland 
}

\begin{document}
\date{Accepted 2021 September 22. Received 2021 August 30; in original form 2021 June 14.}
\pagerange{\pageref{firstpage}--\pageref{lastpage}} \pubyear{xxxx}
\maketitle

\label{firstpage}

\begin{abstract}
Cosmic rays can penetrate planetary atmospheres driving the formation of prebiotic molecules, which are important for the origin of life. We calculate the Galactic cosmic ray fluxes in the habitable zone of five nearby, well-studied solar-type stars and at the orbits of 2 known exoplanets. We model the propagation of Galactic cosmic rays through the stellar winds using a combined 1.5D stellar wind and 1D cosmic ray transport model.

We find that the habitable zone of 61 Cyg A has comparable Galactic cosmic ray fluxes to present-day Earth values. For the other four systems ($\epsilon$ Eri, $\epsilon$ Ind, $\xi$ Boo B and $\pi^1$ UMa), the fluxes are orders of magnitude smaller than Earth values. Thus, it is unlikely that any as-of-yet undetected Earth-like planets in their habitable zones would receive a higher radiation dose than is received on Earth. $\epsilon\,$Ind$\,$b, a Jupiter-like planet orbiting at $\sim$11\,au, receives higher Galactic cosmic ray fluxes than Earth. We find the suppression of Galactic cosmic rays is influenced by whether diffusion or advection dominates at GeV energies and at distances where the wind has reached its' terminal velocity. For advectively-dominated winds ($\sim$younger systems), varying the astrospheric size influences the suppression significantly. For diffusion-dominated systems ($\sim$older systems) the astrospheric size, and therefore knowledge of the ISM properties, are not very important. This reduces the Galactic cosmic ray flux uncertainties in the habitable zone for diffusion-dominated systems. Whether a system is advection- or diffusion-dominated can be determined from the stellar wind properties.
\end{abstract}


\section{Introduction}
\label{sec:intro}

Galactic cosmic rays ionise and dissociate molecules as they penetrate exoplanetary atmospheres. This can lead to the production of prebiotic molecules \citep{dartnell_2011}. It has also recently been suggested that cosmic rays left an imprint on the helicity of DNA via their secondary particles, which are on-average spin polarised \citep{globus_2020,globus_2021}. Thus, Galactic cosmic rays may have been important for the origin of life on Earth and on exoplanets. A number of fingerprint ions, indicative of ionisation by energetic particles (i.e. cosmic rays and stellar energetic particles) in exoplanetary atmospheres, have recently been identified \citep[such as $\mathrm{H_3O}^+$,][]{helling_2019,barth_2021}. Emission from fingerprint ions may be detectable using transmission spectroscopy with the James Webb Space Telescope \citep[JWST,][]{gardner_2006}. This will place limits on the chemical abundances of fingerprint ions, and thus, on the flux of energetic particles in exoplanetary atmospheres. In order to perform chemical modelling of the impact of Galactic cosmic rays on exoplanetary atmospheres \citep[e.g.,][]{griessmeier_2015, barth_2021} that will be characterised with JWST, it is important to model Galactic cosmic ray propagation for different stellar systems. 

The flux of Galactic cosmic rays reaching an exoplanet's orbit depends on two factors: the properties of the stellar wind and the Galactic cosmic ray spectrum in the interstellar medium (ISM). For individual stellar systems knowledge of the local ISM properties, combined with Lyman-$\alpha$ observations, can be used to constrain the star's mass-loss rate \citep[e.g.][]{wood_2004,wood_2014,wood_2021}. Additionally, the ISM properties surrounding a stellar system, combined with the stellar wind properties, determine the size of the astrosphere. $\gamma-$ray observations indicate that the Galactic cosmic ray spectrum in the ISM is similar throughout the Galaxy \citep[with the exception of the Galactic centre,][]{aharonian_2020}. Thus, the local interstellar spectrum (LIS) of Galactic cosmic rays \citep[constrained by \emph{Voyager 1 \& 2} observations outside of the heliosphere,][]{cummings_2016,stone_2019} is often adopted as the spectrum for other stellar systems. 

The LIS is suppressed in an energy-dependent way as the cosmic rays travel through a stellar wind. This was first studied in the context of the solar system \citep[e.g.][]{parker_1965}. The suppression of Galactic cosmic rays varies spatially within the solar system and temporally with the magnetic solar cycle \citep{potgieter_2013}. These variations are collectively described as the modulation of Galactic cosmic rays and affect cosmic rays with $\lesssim 30\,$GeV energies in the solar system. 

Recent efforts have focused on the Galactic cosmic ray fluxes, assuming diffusive cosmic ray transport, for the evolving solar wind \citep[relevant for the origin of life on Earth,][]{rodgers-lee_2020b} and for a number of nearby M dwarf systems \citep{herbst_2020,mesquita_2021} because exoplanets orbiting M dwarf are prime targets in the search for life in the Universe. However, rocky planets in the habitable zone of K and G dwarfs, more similar to our own Sun, are still of great interest.

In addition to Galactic cosmic rays, stellar energetic particles (also known as stellar cosmic rays) are another type of energetic particle that are important for exoplanetary atmospheres \citep{airapetian_2016,tabataba-vakili_2016,chen_2020,rodgers-lee_2021}. Stellar energetic particle fluxes should be largest at young ages when the star's protoplanetary disk is still present \citep[e.g.][]{rab_2017,rodgers-lee_2017,rodgers-lee_2020}. However, here we only focus on the Galactic cosmic ray fluxes.

In this paper, we study the modulation of Galactic cosmic rays for a sample of solar-type stars (K and G dwarfs). The sample comprises of 5 nearby stellar systems ($<14.4\,$pc away) that have both stellar magnetic field measurements and mass-loss rates derived from Lyman-$\alpha$ observations, that we use to construct well-constrained stellar wind models. We use the stellar wind properties as an input for the cosmic ray propagation model. \citet{mesquita_2021b} perform a complementary study considering nearby M dwarfs.

The paper is structured as follows: Section\,\ref{sec:sample} introduces the sample of stars that we consider, Sections\,\ref{sec:form} and \ref{sec:cosmic_ray} briefly discuss the stellar wind and cosmic ray tranport models that are used. Section\,\ref{sec:results} discusses our results and finally, our conclusions are presented in Section\,\ref{sec:conclusions}.

\section{Description of sample}
\label{sec:sample}

Our sample consists of G and K dwarf stars that have a stellar wind mass-loss rate measurement (derived from Lyman-$\alpha$ observations) and ISM velocities \citep[from][and references therein]{wood_2014}, in addition to an observed X-ray luminosity \citep[from][]{vidotto_2014,vidotto_2016} and a stellar magnetic field map \citep[from][]{vidotto_2014,boro-saikia_2016,vidotto_2016}. These observed properties are important to construct well-constrained stellar wind models for individual stars which is discussed in Section\,\ref{sec:form}. The resulting sample consists of 5 nearby stars: 61 Cyg A, $\epsilon$ Eri, $\epsilon$ Ind, $\pi^1$ UMa and $\xi$ Boo B. Four of the stars are within $\sim 7\,$pc of Earth placing them close to, or within, the local interstellar cloud \citep{redfield_2000}. $\pi^1$ UMa is slightly further away ($\sim$14.4\,pc). The stellar masses range from 0.66 - 1$M_\odot$ and the stellar rotation rates vary from 0.8 - 5.4\,$\Omega_\odot$. Age estimates for the systems range from 0.4 to 6\,Gyr. The stellar masses, radii and rotation rates for our sample are from \citet{vidotto_2014,vidotto_2016} and \citet{feng_2019}. The average unsigned radial stellar surface magnetic field strength values range from 3.8 - 18\,G from magnetic maps presented in \citet{vidotto_2014,vidotto_2016} and \citet{boro-saikia_2016} using the Zeeman Doppler Imaging (ZDI) technique\footnote{The years of the observations for 61 Cyg A, $\epsilon$ Eri, $\epsilon$ Ind, $\pi^1$ UMa and $\xi$ Boo B are 2010, 2008, 2013, 2012 and 2007, respectively.}. \citet{vidotto_2014,vidotto_2016} and \citet{boro-saikia_2016} present the average squared stellar magnetic field strength $\langle B^2\rangle =\frac{1}{4\pi}\int(B_r^2+B_\theta^2+B_\varphi^2)\sin\theta d\theta d\varphi $ or $\langle |B|\rangle$. We use $\langle |B_r|\rangle$ for our stellar wind model since stellar winds are thought to be launched from open magnetic field lines. The observationally constrained stellar properties are given in Table\,\ref{table:stellar_parameters}.

Two of the stars are known to host Jupiter-like planets, namely $\epsilon$ Eri b \citep[$1.56M_\mathrm{J}$ orbiting at 3.39\,au,][]{hatzes_2000,benedict_2006, mawet_2019} and $\epsilon$ Ind b \citep[$3.25M_\mathrm{J}$ orbiting at 11.55\,au,][]{feng_2019}, detected via radial velocity observations. Recent high-contrast imaging at 10$\mu$m  by \citet{pathak_2021} with VISIR were not able to directly detect these planets. $\epsilon$ Eri is still surrounded by a debris disk with an estimated age of 0.4\,Gyr. They estimate that, had $\epsilon$ Ind been 0.7\,Gyr old, they would have detected the planet and that their non-detection implies an older age for $\epsilon$ Ind \citep[consistent with other age estimates of 3.7-5.7\,Gyr,][]{feng_2019}. Additionally, it is possible that there are undetected rocky Earth-like planets residing in the habitable zones of the stars in our sample. 

$\pi^1$ UMa is a very interesting object as it is considered to be a young solar analogue (0.5\,Gyr, $1R_\odot, 1\,M_\odot$). In contrast to predictions from spin down models \citep{johnstone_2015b}, \citet{wood_2014} inferred a less-than-solar mass-loss rate ($\dot M$) for $\pi^1$ UMa from Lyman-$\alpha$ observations. \citet{wood_2014} discuss how it could be possible that a higher mass-loss ($150\,\dot M_\odot$) for $\pi^1$ UMa is consistent with the observations if the ISM properties were significantly different (i.e. more ionised and lower in density). $150\,\dot M_\odot$ represents the mass-loss rate obtained if the relation between solar flares and coronal mass ejections is extrapolated to a star as young/active as $\pi^1$ UMa \citep{drake_2013}. This value should be viewed as an upper limit as it is unlikely that $\pi^1$ UMa would have such a high mass-loss rate due to energetic arguments \citep{drake_2013}. Given the discussion surrounding the mass-loss rate for $\pi^1$ UMa we investigate both scenarios: 0.5$\dot M_\odot$ \citep[assuming $n_\mathrm{ISM}(\mathrm{H}\,\textsc{i})=0.16\,\mathrm{cm^{-3}}$ and $n_\mathrm{ISM}(\mathrm{H}^+)=0.08\,\mathrm{cm^{-3}}$ from][]{wood_2014} and $150\,\dot M_\odot$ \citep[assuming $n_\mathrm{ISM}(\mathrm{H}\,\textsc{i})=0.0024\mathrm{cm^{-3}}$ and $n_\mathrm{ISM}(\mathrm{H}^+)=0.0456\mathrm{cm^{-3}}$ from][]{wood_2014}. See \citet{vidotto_2021} for a discussion of the mass-loss rate measurements for $\pi^1$ UMa.

\section{Stellar wind model}
\label{sec:form}

The stellar wind is modelled using solutions of the 1.5D analytic isothermal model of \citet{weber_1967}, presented in \citet{johnstone_2017}. This model accounts for stellar rotation. The wind is launched from the hot corona where it overcomes the star's gravity. The stellar wind accelerates rapidly before reaching its terminal velocity. The stellar properties relevant for the model are the stellar mass ($M_\star$), radius ($R_\star$), rotation rate ($\Omega_\star$), magnetic field strength ($B_\star$), $\dot M$ and temperature of the wind ($T_\mathrm{base}$). $\dot M$ is the only parameter that is varied for each of the stars in our sample. All of the inputs for the stellar wind model are constrained by observations and are shown in Table\,\ref{table:stellar_parameters}.

The wind temperature is assumed to be the temperature at the base of the wind since an isothermal model is used. The coronal temperature, $T_\mathrm{cor}$, can be estimated from stellar X-ray emission as $T_\mathrm{cor} = 0.11\, F_\mathrm{X}^{0.26}$ \citep{johnstone_2015c}. $F_\mathrm{X}$ is the stellar x-ray flux ($\mathrm{erg\,cm^{-2}\,s^{-1}}$) where $F_\mathrm{X} = L_\mathrm{X}/4\pi R_\star^2$. The x-ray luminosities, $L_\mathrm{X}$, for the sample are from \citet{vidotto_2014,vidotto_2016} and are given here in Table\,\ref{table:stellar_parameters}. \citet{ofionnagain_2018} relate coronal temperatures to base temperatures, $T_\mathrm{base}$, by dividing coronal temperatures by a factor of 1.36 to match solar observations near the base of the wind and we follow a similar approach here. Our adopted wind temperatures are shown in Table\,\ref{table:my_parameters}.

\setlength{\tabcolsep}{2pt}
\begin{table*}
\centering
\caption{The stellar properties for the five stars considered are listed here. The columns are: the name of the star, other identifiers for the stars, spectral type, age, mass, radius, rotation period, rotation rate in terms of the present-day solar value ($\Omega_\odot = 2.67\times 10^{-6}\,\mathrm{rad\,s^{-1}}$), x-ray luminosity, average unsigned radial surface magnetic field strength, mass-loss rate from \citet{wood_2002, wood_2005, wood_2014} in terms of the present-day value ($\dot M_\odot = 2 \times 10^{-14}\mathrm{M_\odot\,yr^{-1}}$), ISM velocity, distance to each star and whether an exoplanet has been detected for each system, respectively. References for each property are also given.
}
\begin{tabular}{@{}lllcccccccccccccccc@{}}
\hline

Star & Other identifier & Spec. type & $t_\star$ & $M_\star$ & $R_\star$ & $P_\mathrm{rot}$ & $\Omega_\star$ & log $L_\mathrm{X}$ & $|B_\star|$ & $\dot M_\mathrm{W}$ &   $v_\mathrm{ISM}$ & $d$ & Known \\
&&&&&&&&&&&&& Exoplanet? \\
\hline
 &  & & [Gyr] & [$M_\odot$] & [$R_\odot$] & [day] & [$\Omega_\odot$] & $[\mathrm{erg\,s^{-1}}]$ & [G] &[$\dot M_\odot$] & $\mathrm{[km\,s^{-1}]}$ & [pc] & &\\
\hline

61 Cyg A      & HD 201091A& K5 V$^a$  & 3.6$^a$ & 0.66$^a$ & 0.62$^a$ & 34.2$^a$ & 0.80 & 28.22$^a$ & 3.8$^{c}$ & 0.5$^b$  & 86$^b$ & 3.5$^b$ & No \\
$\epsilon$ Eri& HD 22049  & K2 V$^a$  & 0.4-0.8$^{a,d,e}$  & 0.86$^a$ & 0.77$^a$ & 10.3$^a$ & 2.6 & 28.32$^a$ & 8.1$^a$ & 30.0$^b$ & 27$^b$ & 3.2$^b$ & Yes \\ 
$\epsilon$ Ind& HD 209100 & K5 V$^f$   & 3.7-5.7$^g$ & 0.75$^g$ & 0.75$^f$ & 37.2$^f$ & 0.73 & 27.39$^f$& 17.6$^f$   & 0.5$^b$   & 68$^b$ & 3.6$^b$  & Yes \\
$\pi^1$ UMa   & HD 72905  & G1.5 V$^a$ & 0.5$^a$	& 1.00$^a$ & 1.00$^a$ & 5.00$^a$ & 5.4  & 28.97$^a$ & 7.6$^a$ & 0.5$^h$ & 34$^h$ & 14$^h$ & No \\  

$\xi$ Boo B   & HD 131156B& K4V$^a$   &	2$^a$	&   0.99$^a$ & 1.07$^a$ 	  & 10.3$^a$ & 2.64 & 27.97$^a$  & 11.8$^a$ & 4.5$^i$  & 32$^i$ 	 	& 6.70$^i$ & No \\
\hline 

\end{tabular}

(a) \citet{vidotto_2014}; (b) \citet{wood_2002}; (c) \citet{boro-saikia_2016}; (d) \citet{mamajek_2008}; (e) \citet{mawet_2019}; (f) \citet{vidotto_2016}; (g) \citet{feng_2019}; (h) \citet{wood_2014}; (i) \citet{wood_2005}.\\

\label{table:stellar_parameters}
\end{table*}

Stellar winds from low-mass stars are generally very challenging to measure because of the low densities of the winds. \citet{wood_2002,wood_2005,wood_2014} derive mass-loss rates for the stars in our sample from Lyman-$\alpha$ observations which constrain the total neutral hydrogen column density between us, the observer, and the star. Lyman-$\alpha$ emission from the star is absorbed by neutral hydrogen at the edge of the astrosphere, the ISM and the edge of the heliosphere before being observed at Earth. The ISM ram pressure ($P_\mathrm{ISM}$), and therefore the stellar wind ram pressure ($P_\mathrm{ram}$), most compatible with the observations is then inferred using typical values for the ISM density surrounding the stellar system and the ISM velocity. $P_\mathrm{ram}$ depends on the stellar wind velocity ($v$) and $\dot M$, i.e. $P_\mathrm{ram} = \rho v^2 = \frac{\dot Mv}{4\pi r^2}$ using $\dot M = 4\pi r^2\rho v$, where $\rho$ is the stellar wind density. In \citet{wood_2002,wood_2005,wood_2014}, a solar wind terminal velocity ($v_{\infty,\odot}$) is used, i.e. 400$\,\mathrm{km\,s^{-1}}$, to calculate $\dot M$ for all of the stars. 

The mass loss-rates from \citet{wood_2002,wood_2005,wood_2014} are denoted by $\dot M_\mathrm{W}$ and are shown in Table\,\ref{table:stellar_parameters} for clarity. However, the stellar wind terminal velocities for low-mass stars should vary from star to star due to different coronal temperatures, stellar masses and radii \citep{parker_1965, weber_1967}. We adopt a different approach to \citet{wood_2002,wood_2005,wood_2014} and drop the assumption that $v_{\infty}=v_{\infty,\odot}$ for all of the stars. We use $\dot M_\mathrm{W}$ and $v_{\infty,\odot}$ to calculate $P_\mathrm{ram}$ at a given radius where the wind has reached its terminal velocity for each star. We then use the terminal velocities obtained from the analytic Weber-Davis stellar wind model and vary the mass-loss rate in the model ($\dot M_\mathrm{WD}$) until we match $P_\mathrm{ram}$ at the same given radius \citep[which is the same methodology as adopted by][]{holzwarth_2007}. The terminal wind velocities obtained for our sample of stars are larger than $v_{\infty,\odot}=400\,\mathrm{km\,s^{-1}}$, as adopted by \citet{wood_2002,wood_2005,wood_2014}. Thus, the values of $\dot M_\mathrm{WD}$ that we find are correspondingly lower (given in Table\,\ref{table:my_parameters}) in order to match $P_\mathrm{ram}$.

The magnetic field and velocity of the stellar winds as a function of orbital distance are used in the cosmic ray transport model which is described in Section\,\ref{sec:cosmic_ray}. The wind profiles used are shown in Figure\,\ref{fig:profiles}, as well as a brief description of how the stellar wind model outputs are extrapolated for the cosmic ray model. $P_\mathrm{ram}$ is used to calculate the size of the astrosphere for each star in Section\,\ref{subsec:Rast}. For comparison, we also model the solar wind. For this model $|B_\odot|=1.9\,$G \citep[solar minimum value from][]{vidotto_2014} and $T_\mathrm{base,\odot}=1.5$\,MK \citep{ofionnagain_2018} is used resulting in $v_\infty \sim 560\,\mathrm{km\,s^{-1}}$. This value is larger than the value adopted by \citet{wood_2002, wood_2005, wood_2014} but both are consistent with observations of the solar wind \citep{mccomas_2008}.

\subsection{Astrospheric size}
\label{subsec:Rast}
The size of a star's astrosphere, $R_\mathrm{ast}$, represents the star's dynamic sphere of influence and is determined by the balance of $P_\mathrm{ram}$ and $P_\mathrm{ISM}$. This can be expressed as
\begin{equation}
R_\text{ast} = \left(\frac{P_\text{ram}(r)}{P_\mathrm{ISM}}\right)^{1/2}r
\end{equation}
where $r$ is a given radius where the wind has reached its terminal velocity. $P_\mathrm{ISM} = mn_\mathrm{ISM}v_\text{ISM}^2$, with $n_\mathrm{ISM}$ and $v_\text{ISM}$ being the total number density and velocity of the ISM adjacent to the stellar system. The proton mass is denoted by $m$. We use the same ISM neutral hydrogen density ($n_\mathrm{ISM}(\mathrm{H}\,\textsc{i})=0.14\,\mathrm{cm^{-3}}$) and proton density ($n_\mathrm{ISM}(\mathrm{H}^+)=0.1\,\mathrm{cm^{-3}}$) as given by Model 10 in Table\,1 of \citet{wood_2000} that are relevant for nearby systems in the local ISM to derive the total ISM number density\footnote{For $\pi^1$ UMa, the values given in Section\,\ref{sec:sample} are used \citep[from][]{wood_2014}.}. The ISM velocity that each star sees represents the combined stellar motion and local interstellar cloud flow. The same observationally inferred ISM velocities are used for our sample as quoted in \citet[][and references therein]{wood_2002,wood_2005,wood_2014}, given in our Table\,\ref{table:stellar_parameters}. The astrospheric radii that we derive are given in Table\,\ref{table:my_parameters}. The variation in astrospheric sizes is quite large, ranging from 17 - 2760\,au. In comparison, the heliosphere is approximately 122\,au in radius \citep[as observed by \emph{Voyager 1},][]{krimigis_2013,stone_2019}. $R_\mathrm{ast}$ is used as a fixed outer spatial boundary condition for the cosmic ray transport model.

\begin{table}
\centering
\caption{Stellar wind properties and other parameters relevant for the cosmic ray simulations. The columns are: the star's name, the stellar wind terminal velocity, the mass-loss rate obtained using the Weber-Davis model, the astrospheric size, the ISM ram pressure, the habitable zone (HZ) location and the wind base temperature, respectively.  }
\begin{tabular}{@{}lcccccccccccccccccc@{}}
\hline

Star & $v_\infty$& $\dot M_\mathrm{WD}$ & $R_\mathrm{ast}$ & $P_\mathrm{ISM}$ & HZ & $T_\mathrm{base}$   \\
\hline
  & [$\mathrm{km\,s^{-1}}$]& [$\dot M_\odot \mathrm{yr^{-1}}$] & [au] & $\mathrm{[g\,cm^{-1}\,s^{-2}]}$& [au] & [MK]  \\
\hline

61 Cyg A       &840  & 0.2 & 17  & $3\times10^{-11}$ & 0.3-0.8  & 2.7  \\
$\epsilon$ Eri &910	 & 15  & 430 & $3\times10^{-12}$ & 0.5-1.1  & 2.6  \\
$\epsilon$ Ind &580	 & 0.3 & 22  & $2\times10^{-11}$ & 0.4-0.9  & 1.5  \\
$\xi$ Boo B    &770  & 2.7 & 140 & 		$4\times10^{-12}$			   &	0.5 - 1.3		&   1.7   \\
$\pi^1$ UMa    &1380 & 0.2 & 44  & $5\times10^{-12}$ & 0.8-1.8   & 3.3  \\  
\hline
$\pi^1$ UMa    &1050 & 150 & 2760 & $5\times10^{-12}$ & 0.8-1.8   & 3.3 \\  
\hline
\label{table:my_parameters}
\end{tabular}
\end{table}

\section{Cosmic ray model}
\label{sec:cosmic_ray}
We use a 1D diffusive-advection transport equation for the propagation of the Galactic cosmic rays. The model is described in detail in \citet{rodgers-lee_2020b}. However, we include a brief description here for completeness. \citet{parker_1965} used the diffusive-advection transport equation in the context of the modulation of Galactic cosmic rays in the solar system. The 1D transport equation is given by
\begin{equation}
\frac{\partial f}{\partial t} = \nabla\cdot(\kappa\nabla f)-v\cdot\nabla f +\frac{1}{3}(\nabla\cdot v)\frac{\partial f}{\partial \mathrm{ln}p}
\label{eq:f}
\end{equation}
\noindent where $f$, $\kappa(p,B(r))$ and $p$ are the cosmic ray phase space density, spatial diffusion coefficient and momentum, respectively. $v(r)$ is the stellar wind velocity obtained from the stellar wind model. The first term on the righthand side of Eq.\,(\ref{eq:f}) accounts for the spatial diffusion of Galactic cosmic rays into the stellar system. The second and third terms act to suppress the flux of Galactic cosmic rays in different ways. The second term represents how the stellar wind advects the cosmic rays back out of the system, opposing their inward diffusion. The third term represents adiabatic losses due to the expansion of the stellar wind. This shifts the cosmic rays to lower energies. Eq.\,\ref{eq:f} does not account for cosmic ray drift motions \citep{jokipii_1977}. As noted in \citet{rodgers-lee_2020b} this means that latitudinal variations and temporal variation of Galactic cosmic ray modulation due to a stellar activity cycle cannot be studied here. 2D or 3D cosmic ray transport models \citep[such as][]{cohen_2012,potgieter_2015} are needed to include these effects.

The diffusion coefficient, $\kappa$, for the cosmic rays depends on the level of turbulence in the magnetic field and on their momentum. Based on quasi-linear theory \citep{jokipii-1966,schlickeiser_1989}, $\kappa$ can be expressed as
\begin{equation}
\frac{\kappa(p,B(r))}{\beta c} = \eta_0\left( \frac{p}{p_0} \right) ^{1-\gamma} r_\mathrm{L}
\label{eq:kappa}
\end{equation}
\noindent where $r_\text{L}$ is the Larmor radius of the cosmic rays and $p_0=3$\,GeV$/c$. Eq.\,\ref{eq:kappa} assumes isotropic diffusion. In reality, there are different components for parallel, $\kappa_\parallel$, and perpendicular, $\kappa_\perp$, diffusion with respect to the magnetic field direction. Similar to \citet{herbst_2020}, as a first approximation we have assumed in Eq.\,\ref{eq:kappa} that $\kappa = \beta \lambda/3 $ where the particle mean free path, $\lambda$, scales with $r_\text{L}$. A more complete description of the diffusion coefficient requires more information than is currently available about the type of turbulence present in these systems. The power law index $\gamma$ represents the type of turbulence present, such as Bohm diffusion ($\gamma=1$), Kolmogorov-type turbulence ($\gamma=5/3$) or magnetohydrodynamic-driven turbulence ($\gamma=3/2$).  Here we adopt $\gamma=1$, similar to \citet{rodgers-lee_2020b}. $\eta_0$ relates to the level of turbulence in the magnetic field \citep[see Eq.(4) from][]{rodgers-lee_2020b}. For the cosmic ray simulation of the solar system we use $R_\mathrm{ast}=122\,$au (available from \emph{Voyager} observations) and $\eta_0=1$. There are currently no observational constraints for the type or level of turbulence present in other stellar systems. Constraining the chemical abundances of fingerprint ions in exoplanetary atmospheres could provide constraints on the incident cosmic ray spectrum. This can potentially be used to infer the type/level of turbulence present. In this case, the exoplanetary atmosphere and host star properties need to be well-constrained. It is possible that there is an increased level of turbulence in young systems due to increased stellar activity leading to more coronal mass ejections (CMEs). CMEs are thought to drive turbulence in the solar wind \citep{cranmer_2017}. However, simulations suggest that the strong magnetic fields of young stars may confine CMEs \citep{alvarado-gomez_2018}. Overall, it is unclear how the values of $\gamma$ and $\eta_0$ would change for systems of different ages. We adopt $\gamma=1$ and $\eta_0=1$ for all of our simulations for simplicity \citep[similar to][]{svensmark_2006,cohen_2012}. It is important to note that adopting different values for $\gamma$ and $\eta_0$ would affect the results presented here.

For the boundary conditions, the inner spatial boundary condition is reflective. The outer spatial boundary condition is a fixed boundary condition which is set to be the LIS values using the model fit to the \emph{Voyager 1} data given in Eq.\,(1) of \citet{vos_2015}. The inner and outer momentum boundary conditions are both outflow boundary conditions. The spatial and momentum grids are logarithmically spaced. The spatial grid ranges from 0.01\,au to $R_\mathrm{ast}$ for each star (given in Table\,\ref{table:my_parameters}). The momentum grid ranges from 0.15 - 100\, GeV$/c$ with the number of momentum bins, $N_\mathrm{p}=60$.

\section{Results}
\label{sec:results}

\subsection{Galactic cosmic ray intensities in the habitable zones}
\begin{figure*}
	\centering
        \includegraphics[width=\textwidth]{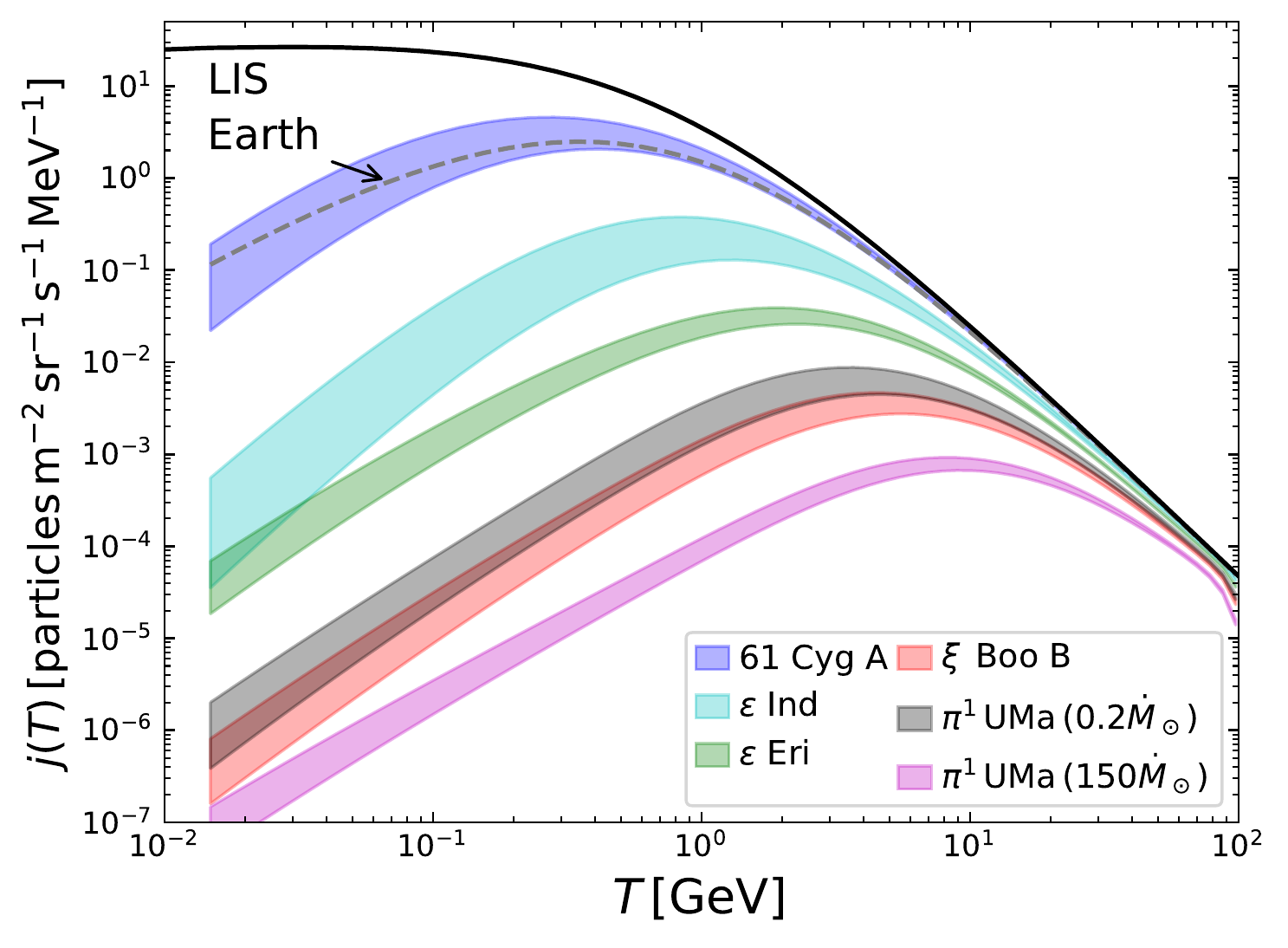}
       	\centering
  \caption{The differential intensities of Galactic cosmic rays in the habitable zone are shown by the shaded regions for each star in the sample as a function of cosmic ray kinetic energy. The solid black line represents the LIS. The grey dashed line shows the differential intensities at Earth for comparison.\label{fig:hzs}} 
\end{figure*}

The flux of Galactic cosmic rays in the habitable zone (shaded regions) for our sample is shown as a function of cosmic ray kinetic energy in Fig.\,\ref{fig:hzs}. The habitable zones for the stars are calculated using Eq.\,(4) and (5) from \citet{kopparapu_2014} with the recent Venus and early Mars criteria valid for planetary masses between $0.1M_\oplus \leq M_p \leq 5 M_\oplus $ (the values are given in Table\,\ref{table:my_parameters}). The grey dashed line indicates typical Galactic cosmic ray fluxes at Earth for comparison. The solid black line shows the LIS fit which is the fixed outer boundary condition. Fig.\,\ref{fig:hzs} includes the two different scenarios for $\pi^1$ UMa, assuming $\dot M_\mathrm{W}=0.5 \dot M_\odot$ (corresponding to $\dot M_\mathrm{WD}=0.2\dot M_\odot$, grey shaded region) inferred from the observations of \citet{wood_2014} and $M_\mathrm{WD}=150\dot M_\odot$ (magenta shaded region) based on the extrapolation of the relation between solar flares and CMEs from \citet{drake_2013}. 

61 Cyg A is the only system with similar Galactic cosmic ray fluxes in the habitable zone as observed at Earth (by comparing the blue shaded region and grey dashed line in Fig.\,\ref{fig:hzs}). The four other systems have much lower Galactic cosmic ray fluxes in their habitable zones. At GeV energies, there are approximately 6 orders of magnitude difference at GeV cosmic ray energies between the two most extreme systems (61 Cyg A and $\pi^1$ UMa assuming $\dot M_\mathrm{WD} = 150 \dot M_\odot $). For $\pi^1$ UMa, the main result of adopting a different $\dot M$ (along with different ISM densities) for the cosmic ray transport model is the astrospheric size. For $\dot M_\mathrm{WD} = 0.2\dot M_\odot$, $R_\text{ast} = 44$\,au, whereas for $\dot M_\mathrm{WD} = 150 \dot M_\odot $, then $R_\text{ast} = 2760$\,au, i.e. 63 times bigger. At GeV energies, this changes the Galactic cosmic ray fluxes in the habitable zone by approximately 2 orders of magnitude (the grey and magenta shaded regions in Fig.\,\ref{fig:hzs}). The kinetic energy at which the peak cosmic ray flux occurs also shifts from 4\,GeV to 8\,GeV. Thus, adopting a different $\dot M$ and ISM densities for $\pi^1$ UMa significantly affects the Galactic cosmic ray fluxes in the habitable zone. 

The habitable zones shown in Fig.\,\ref{fig:hzs} are at different orbital distances due to the stars' different spectral types and luminosities (shown in Table\,\ref{table:my_parameters}). We compared the Galactic cosmic ray fluxes at 1\,au for the different systems. The general behaviour was very similar to what is seen in Fig.\,\ref{fig:hzs} with 61 Cyg A having the largest fluxes and $\pi^1$ UMa showing the smallest. However, this behaviour would not necessarily be expected at all orbital distances.

\begin{figure*}%
	\centering
    \subfigure[61 Cyg A]{%
        \includegraphics[width=0.5\textwidth]{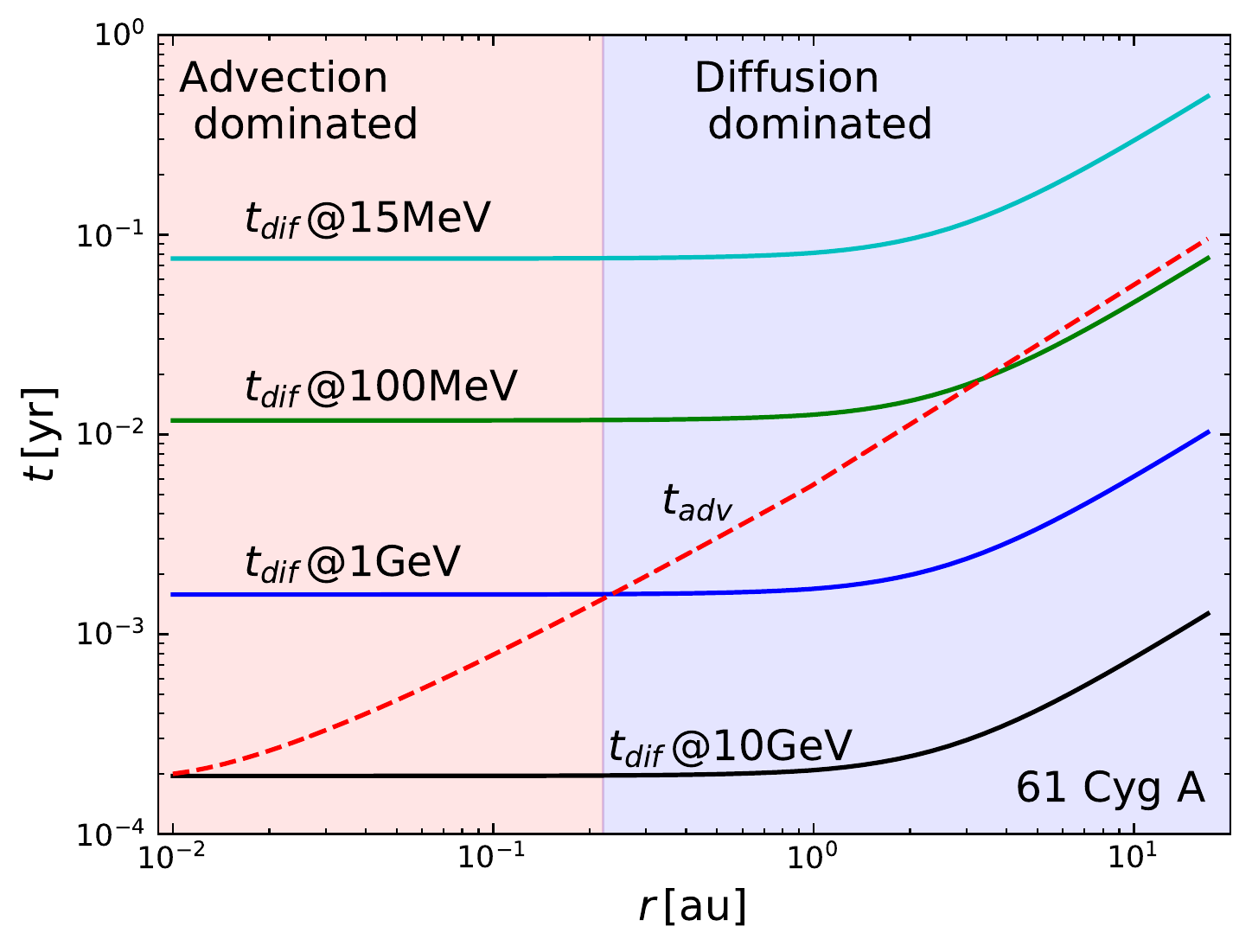}
       	\centering
\label{fig:tdiff_17au}}%
~
	\centering
    \subfigure[$\pi^1$ UMa ]{%
        \includegraphics[width=0.5\textwidth]{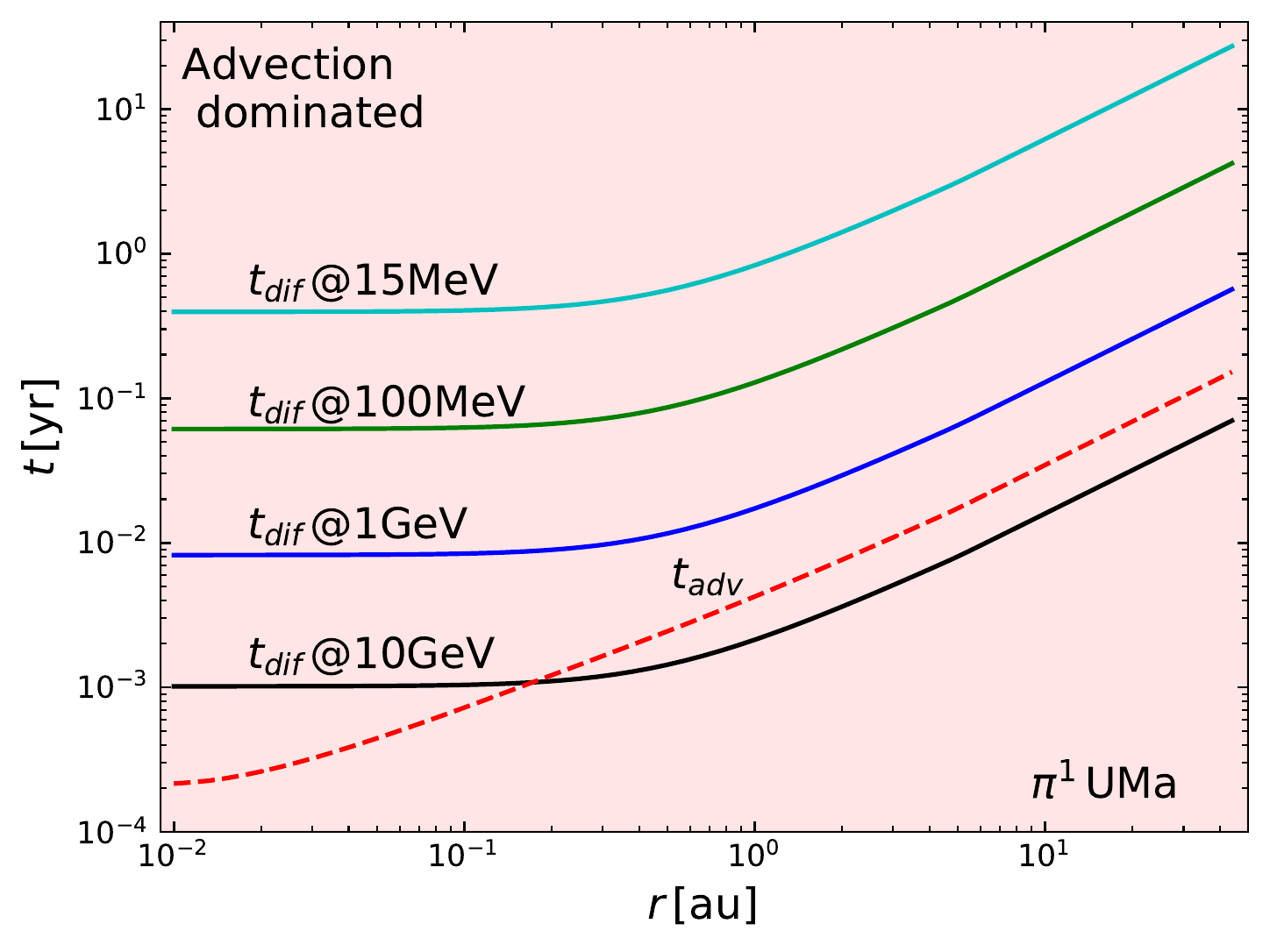}
       	\centering
\label{fig:tdiff_44au}}%
    \caption{The diffusion timescales for different cosmic ray energies (solid lines) and the advective timescale (red dashed line) as a function of orbital distance are shown for two systems representative of the sample that we consider. The blue shaded region in (a) for the 61 Cyg A system represents a ``diffusion-dominated'' region where $t_\mathrm{dif}<t_\mathrm{adv}$ for GeV energy cosmic rays. The red shaded regions in (a) and (b) represent ``advection-dominated'' regions where $t_\mathrm{dif}>t_\mathrm{adv}$ for GeV energy cosmic rays.} 
    \label{fig:timescales}%
\end{figure*}

In order to understand the level of modulation suffered by the Galactic cosmic rays for the different systems it is necessary to consider two factors, namely the timescales for the different physical processes and the astrospheric size. Individual stellar properties, such as $B_\star$ or $\Omega_\star$ for instance, cannot be used directly to understand the Galactic cosmic ray modulation for the systems. The diffusive and advective timescales are given by
\begin{equation}
t_\mathrm{dif} = \frac{r^2}{\kappa(p,B(r))},\hspace{5mm} t_\mathrm{adv} = \frac{r}{v(r)}.
\end{equation}
If $t_\mathrm{dif}< t_\mathrm{adv}$ for a given cosmic ray energy then diffusion dominates. It means that the cosmic rays of that energy can diffuse into the system faster than the advective processes can act to suppress the Galactic cosmic rays. Fig.\,\ref{fig:timescales} shows the timescales as a function of orbital distance for two systems representative of the sample: 61 Cyg A and $\pi^1$ UMa ($\dot M_\mathrm{WD}=0.2\dot M_\odot$ case). The diffusion timescale depends on momentum and so timescales for different cosmic ray energies are shown. 

We find that the ratio of $t_\mathrm{dif}$ for GeV energy cosmic rays (solid blue line) versus $t_\mathrm{adv}$ (dashed red line) at large orbital distances, where the wind has reached its' terminal velocity and $B_\phi$ dominates, is a good indicator of the overall level of modulation that Galactic cosmic rays will suffer from. At these distances, $t_\mathrm{dif}/t_\mathrm{adv}$ remains constant with orbital distance because $t_\mathrm{dif}\propto r^2B_\phi \propto r$ and $t_\mathrm{adv} \propto r$. 

For example, for 61 Cyg A at 1\,GeV and $r=R_\mathrm{ast}$ in Fig.\,\ref{fig:tdiff_17au}, $t_\mathrm{dif}/t_\mathrm{adv}\sim 0.1$ meaning that diffusion dominates and the cosmic ray fluxes are not greatly suppressed (indicated by the blue shaded region). Advective processes only dominate at $r\lesssim 0.2$\,au for $\lesssim$GeV cosmic ray energies (indicated by the red shaded region). The opposite extreme is $\pi^1$ UMa, where $t_\mathrm{dif}/t_\mathrm{adv}\sim 4$ for GeV energies at $r=R_\mathrm{ast}$ (irrespective of the $\dot M$ adopted since $v_\infty$ and $B(r)$ are very similar for both cases) and the cosmic ray fluxes are significantly suppressed, shown in Fig.\,\ref{fig:tdiff_44au}. Advective processes dominate throughout the wind for $\pi^1$ UMa at $\lesssim$GeV cosmic ray energies (indicated by the red shaded region). This is why for the larger $\dot M$ (or larger astrospheric radius) case, we see significantly more suppression of the Galactic cosmic ray fluxes. If instead diffusion dominated at large orbital distances changing the size of $R_\mathrm{ast}$ would make little difference to the Galactic cosmic ray fluxes in the habitable zone, as found in \citet{mesquita_2021} for the GJ\,436 system.

It is likely that there will be variations from this `rule' that we have proposed here. For instance, as we have discussed the size of $R_\mathrm{ast}$ for advection-dominated systems affects the modulation of Galactic cosmic rays. $R_\mathrm{ast}$ depends on the ISM properties. This is not a property intrinsic to the star itself and thus, variation from star to star should be expected. Another property that may lead to deviations from our rule is $\Omega_\star$. We are relying on the value of $t_\mathrm{dif}/t_\mathrm{adv}$ at large orbital distances to guide us, i.e. where the wind has reached its' terminal velocity and therefore the ratio $t_\mathrm{dif}/t_\mathrm{adv}$ is constant with orbital distance. However, at large orbital distances the magnetic field changes from being dominated by the radial component to being dominated by the azimuthal component, i.e. the tightening of the Parker spiral. This occurs for the stars in our sample at approximately 1\,au which is also approximately where the habitable zone for the stars is located. This means that $t_\mathrm{dif}/t_\mathrm{adv}$ also changes in this region, generally becoming more advection-dominated. Thus, the location of this turn-over point will affect the modulation of Galactic cosmic rays \citep[see][for a more detailed discussion of this effect]{mesquita_2021b}.

\subsection{Galactic cosmic ray intensities at planets' orbits}
\begin{figure}
	\centering
        \includegraphics[width=0.5\textwidth]{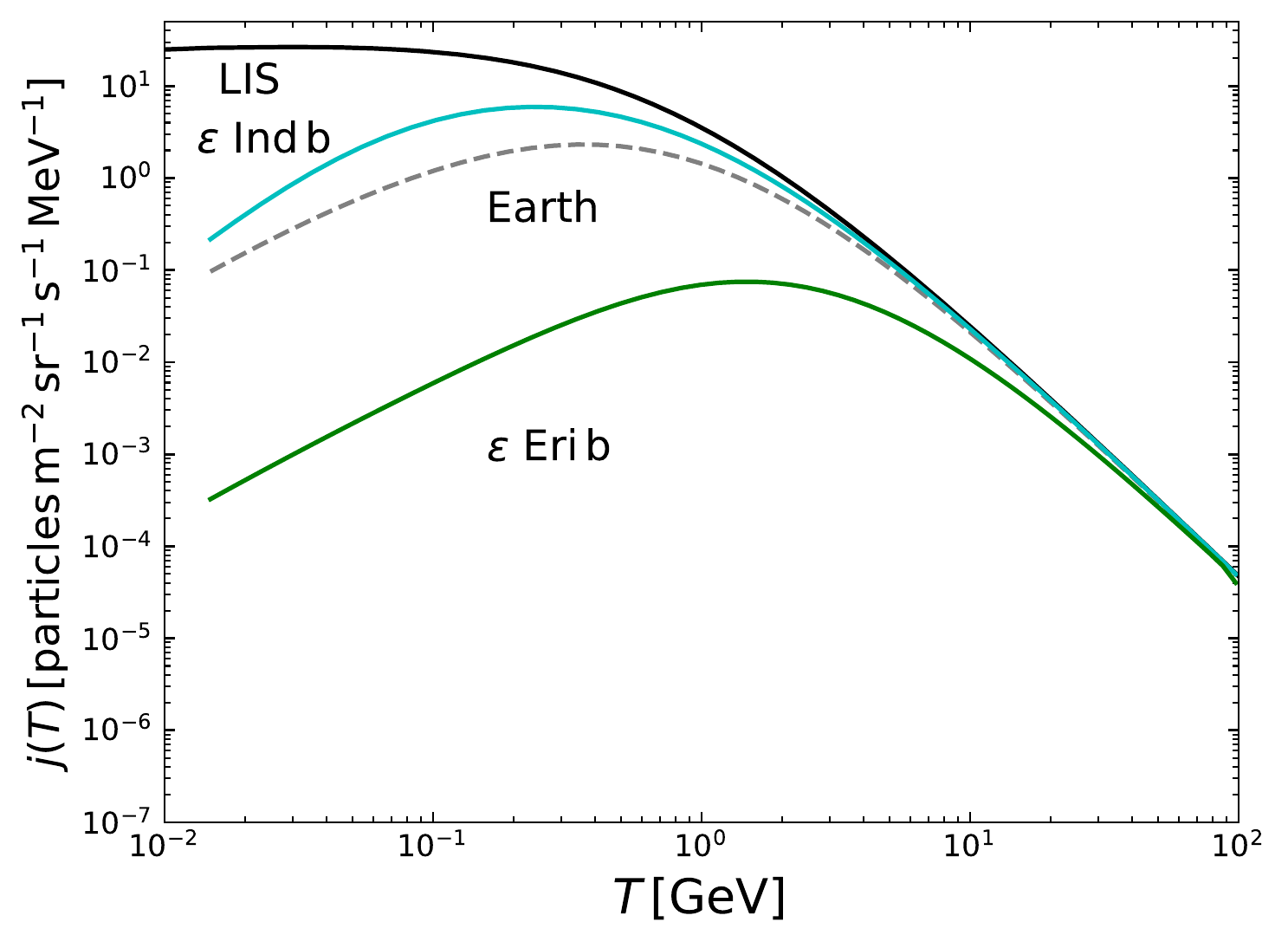}
       	\centering
  \caption{The Galactic cosmic ray fluxes for the planets in our sample, $\epsilon$ Eri\,b (solid green line) and $\epsilon$ Ind\,b (solid cyan line). The solid black line represents the LIS. The grey dashed line shows the differential intensities at Earth for comparison.\label{fig:planets}} 
\end{figure}

For the two systems which host known exoplanets ($\epsilon$ Eri and $\epsilon$ Ind), the Galactic cosmic ray flux at the orbital distance of the planets is shown in Fig.\,\ref{fig:planets}. Again, for comparison we overplot the LIS and fluxes at Earth, denoted by the solid black line and dashed grey line, respectively. For $\epsilon$ Ind b, an approximately $\sim 3$ Jupiter mass planet orbiting at 11.55\,au, it receives larger Galactic cosmic ray fluxes than Earth. On the other hand, the fluxes reaching $\epsilon$ Eri b, an approximately Jupiter mass planet orbiting at 3.39\,au, are much lower. 

\citet{mawet_2019} suggested that $\epsilon$ Eri\,b should be detected via direct imaging with JWST's NIRCam instrument, depending on model assumptions for the exoplanet. They also discussed that spectroscopic age indicators suggest that the star is closer to 800\,Myr old, rather than 200\,Myr. Therefore, they indicated the planet will have an effective temperature of $\sim$150\,K. Spectroscopic observations with JWST's NIRSpec instrument could additionally probe molecular abundances in the upper atmosphere of the planet. Modelling of the chemical impact of Galactic cosmic rays in $\epsilon$ Eri\,b could investigate whether there are any signatures that will be observable with JWST. This could be used to constrain our models in the future. $\epsilon$ Ind is thought to be an older system and therefore it is less likely that the exoplanet would be detectable with JWST \citep{viswanath_2021}.

A number of studies have previously focused on M dwarf systems, some of which host known exoplanets \citep{herbst_2020,mesquita_2021,mesquita_2021b}. \citet{herbst_2020} use 3D magneto-hydrodynamic simulations to model the interaction of the stellar wind with the ISM for 3 M dwarf systems. These simulations give more information about the stellar equivalent of the heliosheath and the termination shock which we do not account for in our model. However, in comparison to \citet{herbst_2020}, local ISM properties were available for all of the stars in our sample which is important for advection-dominated systems. \citet{mesquita_2021} used similar ISM observational constraints for GJ\,436 with an estimate for its stellar magnetic field strength. Thus, our sample represents the most well-constrained sample. \citet{mesquita_2021b} perform a complementary study considering nearby M dwarf systems (using a similar selection criteria as for our sample).

\citet{herbst_2020} found significant Galactic cosmic ray fluxes for Proxima Cen\,b and LHS 1140\,b. In comparison, \citet{mesquita_2021} found much lower fluxes reaching GJ 436\,b (although it is important to note that GJ 436\,b orbits much closer in than the habitable zone). Some of the variety of behaviour found between these two studies can be attributed to different cosmic ray transport properties (i.e. Kolmogorov-type turbulence versus Bohm diffusion). However, despite this, a variety of behaviour is observed due to the different stellar wind properties and astrospheric sizes (which affects advection-dominated systems). Thus, while it is now evident that Galactic cosmic rays are an important ingredient to consider for exoplanet habitability, it will hopefully also be possible with JWST observations of exoplanetary atmospheres to constrain our models. The sample from \citet{mesquita_2021b} includes 8 known exoplanets, presenting an exciting opportunity for upcoming observations. The energetic particle flux at a given height in an exoplanetary atmosphere can be constrained by observations that probe the abundances of fingerprint ions, such as $\mathrm{H_3O}^+$ and $\mathrm{NH_4^+}$ \citep{barth_2021}. In theory, the molecular bands for $\mathrm{H_3O}^+$ are within the spectral range of JWST's NIRSpec instrument \citep[e.g.][]{bourgalais_2020}. Thus, future 3D magneto-hydrodynamic simulations of stellar winds, \citep[such as those presented in][]{herbst_2020}, combined with 2D/3D cosmic ray transport models appear well motivated.

\section{Discussion \& Conclusions}
\label{sec:conclusions}
We investigated the suppression, or modulation, of Galactic cosmic rays for 5 nearby, well-studied solar-type stars. We modelled the propagation of Galactic cosmic rays through the stellar wind using a combined stellar wind and cosmic ray transport model. The stellar wind properties are well constrained as both mass-loss rates inferred from observations, stellar magnetic field measurements and ISM velocities are available in the literature for our sample. This represents the most well-constrained sample presented to date and offers valuable physical insight with regard to the importance of the ISM velocities for certain systems. Two of the stars in the sample host known Jupiter-like exoplanets.

We found that the Galactic cosmic ray fluxes in the habitable zone for four of the five stars were significantly lower than the fluxes received at Earth. 61 Cyg A is the only system with comparable fluxes to Earth in its' habitable zone. 61 Cyg A is an interesting system because it displays a solar-like magnetic cycle \citep{boro-saikia_2016} with polarity reversals in the large-scale magnetic geometry. Here we did not investigate the effect of such a magnetic cycle on the modulation of Galactic cosmic rays, but similar to the solar cycle it would most likely introduce cyclic variations in the Galactic cosmic ray fluxes. The drift of cosmic rays would be interesting to study for 61 Cyg A which would require more complex modelling than presented here. In addition, $\epsilon$ Ind\,b, orbiting at 11.55\,au, received higher fluxes than Earth while the other known exoplanet $\epsilon$ Eri\,b is exposed to much lower fluxes. 

We identify two important regimes displayed by the systems in our study. When the diffusion timescale for GeV energy cosmic rays is shorter than the advective timescale (diffusion-dominated) at large orbital distances we find that (a) the Galactic cosmic rays are least suppressed in these systems and (b) varying the astrospheric size would not change the level of suppression significantly \citep[similar to what was found by][for the M dwarf GJ436]{mesquita_2021}. 61 Cyg A is one such system. On the other hand, when the advection timescale is shorter than the diffusion timescale (advection-dominated) we find the opposite: the Galactic cosmic rays are most suppressed in these systems and varying the astrospheric size also changes the level of suppression significantly. $\pi^1$ UMa is an example of an advection-dominated system. Overall, this means that for any diffusion-dominated system knowledge about the astrospheric size (via the ISM properties) is not hugely important. For these systems having a well-constrained stellar wind model will provide sufficient information to robustly estimate the modulation of Galactic cosmic rays. Generally, diffusion-dominated systems are likely to be older due to slower terminal velocities and weaker stellar magnetic fields. It is important to note that our results depend on the cosmic ray transport properties that we adopt. Different cosmic ray transport properties would likely alter our results. A detection of emission from fingerprint ions, such as $\mathrm{H_3O}^+$ and  $\mathrm{NH_4^+}$, from an exoplanetary atmosphere will constrain the incident energetic particle flux which will in turn could constrain the energetic particle transport properties. However, there will also be some dependence on the exoplanet atmosphere properties, such as the chemical composition. Exoplanets orbiting at different distances can be used to disentangle stellar radiation and stellar energetic particle effects from Galactic cosmic rays.

Our results show a wide range of Galactic cosmic ray fluxes in the habitable zone for the stars in our sample. While there are no known Earth-like planets in our sample we can speculate how Galactic cosmic ray fluxes could impact life on an Earth-like planet in the habitable zones. It is very challenging to detect Earth-like planets orbiting in the habitable zone of solar-type stars. Thus, it is entirely possible that undetected Earth-like planets exist in the habitable zone of our sample of stars. The radiation dose can be used as a measure of how damaging ionising radiation (e.g. cosmic rays or their secondary particles) is for cell tissue \citep[see][for example]{atri_2020,atri_2020b}. \citet{atri_2017} indicated that the total atmospheric column depth is the most important quantity for determining the cosmic ray flux reaching the exoplanet's surface, rather than the chemical composition of the atmosphere, for instance. Thus, we can infer that the radiation dose received at the surface of an Earth-like planet in the habitable zone of the stars in our sample will be comparable, or lower, than the radiation dose received on the surface of Earth \citep[$\sim 0.6\,$mSv$/$day, ][]{atri_2020b}. This is because the Galactic cosmic rays fluxes in the habitable zones are comparable (for 61 Cyg A) or lower than those reaching Earth for our sample. This is based on the assumption that any hypothetical Earth-like planet has a similar atmospheric column density to Earth and a similar planetary magnetic field strength. A stronger planetary magnetic field would prevent higher energy cosmic rays reaching the surface of the Earth-like planet \citep{griessmeier_2015}. Overall, this suggests that life as we know it would not be negatively impacted by the Galactic cosmic ray fluxes that we find for our sample.

\section*{Acknowledgements}
We would like to thank the anonymous referee for helpful comments which improved the manuscript.
 DRL would like to thank Carolina Villarreal-d'Angelo for very helpful discussions which improved the paper. The authors acknowledge funding from the European Research Council (ERC) under the European Union's Horizon 2020 research and innovation programme (grant agreement No 817540, ASTROFLOW). The authors wish to acknowledge the Irish Centre for High-End Computing (ICHEC) for the provision of computational facilities and support. AAV and ALM acknowledge funding from the Provost’s PhD Project award.

\section*{Data Availability}
The output data underlying this article will be available via zenodo.org upon publication.

\appendix
\section{Stellar wind profiles}
\label{appendix:swp}

\begin{figure*}%
	\centering
    \subfigure[]{%
        \includegraphics[width=0.5\textwidth]{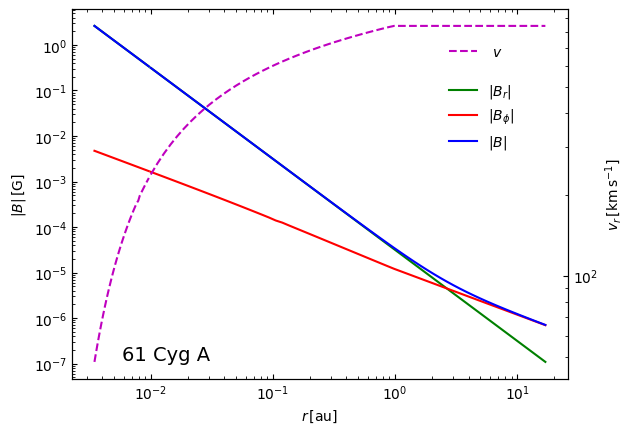}
       	\centering
\label{fig:omega1}}%
  	    ~
    \subfigure[]{%
        \includegraphics[width=0.5\textwidth]{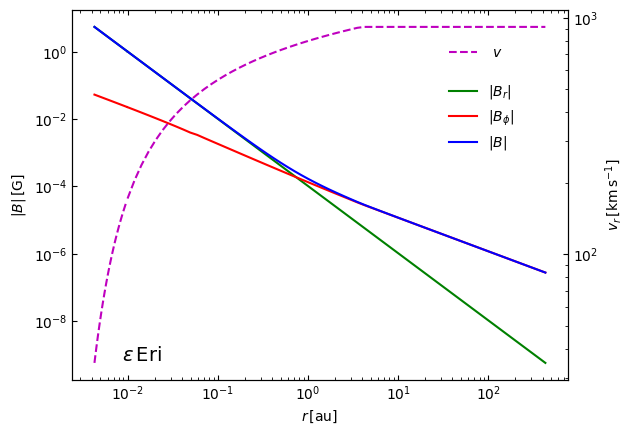}
	\centering
 \label{fig:omega4}}%
\\
	\centering
    \subfigure[]{%
        \includegraphics[width=0.5\textwidth]{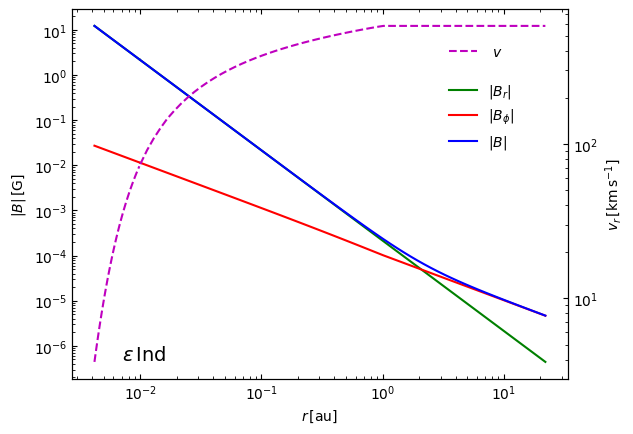}
       	\centering
\label{fig:omega1}}%
  	    ~
    \subfigure[]{%
        \includegraphics[width=0.5\textwidth]{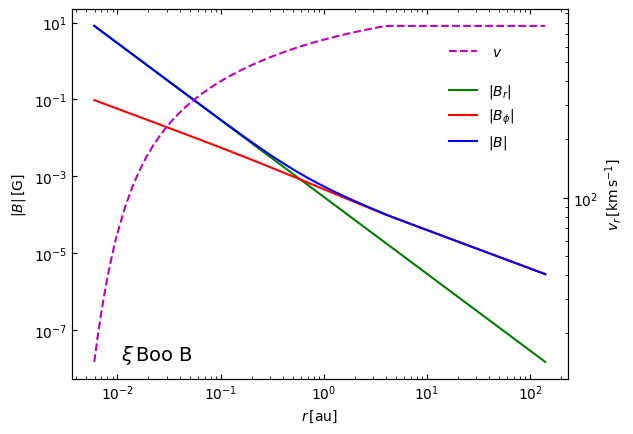}
	\centering
 \label{fig:omega4}}%
\\
	\centering
    \subfigure[]{%
        \includegraphics[width=0.5\textwidth]{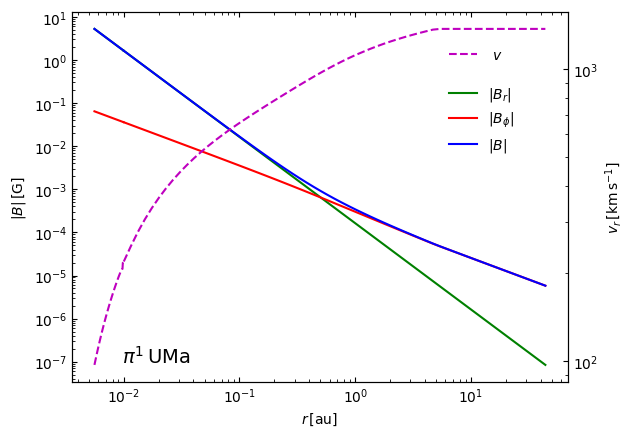}
       	\centering
\label{fig:omega1}}%
  	    ~
    \subfigure[]{%
        \includegraphics[width=0.5\textwidth]{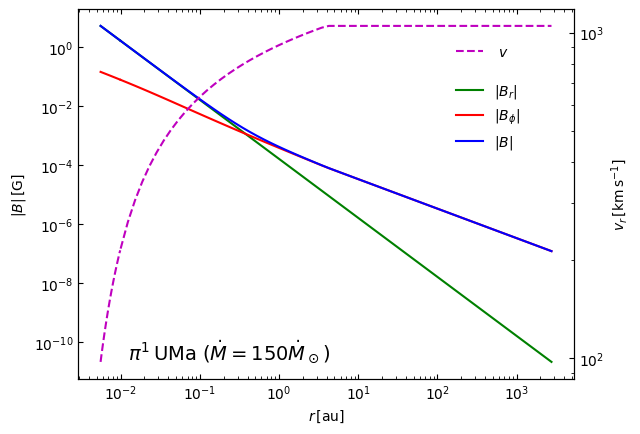}
	\centering
 \label{fig:omega4}}%
    \caption{ The magnetic field and velocity profiles are shown as a function of radial distance for the five stars in our sample. In each panel, the blue, green and red solid lines are the total, radial and azimuthal components of the stellar wind magnetic field, respectively. The dashed magenta line is the velocity profile of the stellar wind. The stellar wind model values extend to $\sim 1\,$au depending on the system, beyond which the data is extrapolated as described in the text. The values used in the cosmic ray simulations start at 0.01\,au.  } 
    \label{fig:profiles}%
\end{figure*}

The stellar wind profiles that are used in the cosmic ray transport simulations for our sample of stars are shown in Fig.\ref{fig:profiles}. The left axis gives the magnetic field strength (total, radial and azimuthal magnetic field strengths shown by solid blue, green and red lines, respectively) and the right axis shows the velocity (dashed magenta lines) as a function of orbital distance. The stellar wind model extends from the base of the wind at $1\,R_\star$ to an orbital distance where the stellar wind has reached its terminal velocity, at 1\,au for the solar system, for instance. Therefore, beyond this distance we extrapolate the stellar wind properties out to the size of each astrosphere for the cosmic ray transport model. The velocity is extrapolated as $v(r> 1\,\mathrm{au})=v(r=1\,\mathrm{au})$. For the magnetic field, $B_r(r> 1\,\mathrm{au})=B_r(r=1\,\mathrm{au})/r^2$ and $B_\phi(r>1\,\mathrm{au})=B_\phi(r=1\,\mathrm{au})/r$.  The inner boundary for the cosmic ray transport model is at 0.01\,au for all of the simulations. The distance at which the wind has reached its terminal velocity is slightly larger than 1\,au for some of the systems considered here which we account for when we extrapolate the data.  

\bibliographystyle{mnras}
\bibliography{../../donnabib}

\label{lastpage}

\end{document}